\documentclass[aps,floatfix,twocolumn,prb,superscriptaddress]{revtex4-2}
\usepackage{amsmath,amssymb}
\usepackage{graphicx}
\usepackage{subfig}
\usepackage{xcolor}
\usepackage{nicefrac}
\usepackage{xspace}

\newcommand{\hamname}{molecular symmetry adapted spin space\xspace}
\newcommand{\hamacro}{mSASS\xspace}

\newcommand{\mat}[1]{\tilde{#1}}
\renewcommand{\vec}[1]{\mathbf{#1}}
\newcommand{\op}[1]{\hat{#1}}
\newcommand{\ket}[1]{\left|#1\right>}

\begin{document}


\title{Coordination chemistry in molecular symmetry adapted spin space (mSASS)}
\author{R. Matthias Geilhufe}
\email{matthias.geilhufe@chalmers.se}
\affiliation{Department of Physics, Chalmers University of Technology, Kemigården 1, 412 58 G\"{o}teborg, Sweden}
\author{Jeffrey D. Rinehart}
\email{jrinehart@ucsd.edu}
\affiliation{Department of Chemistry and Biochemistry, University of California, San Diego, La Jolla, California, USA}

\date{\today}

\begin{abstract} 
Many areas of chemistry are devoted to the challenge of understanding, predicting, and controlling the behavior of strongly localized electrons. Examples include molecular magnetism and luminescence, color centers in crystals, photochemistry and quantum sensing to name but a few. Over the years, an amalgam of powerful quantum chemistry methods, simple intuitive models, and phenomenological parameterizations have been developed, providing increasingly complex and specialized methodologies. Even with increasing specialization, a pervasive challenge remains that is surprisingly universal - the simultaneous description of continuous symmetries (e.g. spin and orbital angular momenta) and discrete symmetries (e.g. crystal field). Modeling and predicting behavior in these complex systems is increasingly important for metal ions of unusual or technologically relevant behavior. Additionally, development and adoption of broad-scope models with physically-meaningful parameters carries the potential to facilitate interdisciplinary collaboration and large-scale meta analysis. Here, we propose a generalized algorithmic approach, the molecular symmetry adapted spin space (mSASS), to localized electronic structure via descent directly from fermionic (spin) rather than bosonic (orbital) symmetry. We derive the Hamiltonian in symmetry-constrained matrix form with an exact account of free parameters and several usage examples. Although preliminary in its implementation, a fundamental benefit of this approach is the treatment of spatial and spin-orbit symmetries without the need for perturbative approximations. In general, the \hamacro Hamiltonian is large but finite and can be diagonalized numerically with high efficiency, providing a basis for conceptual models of electronic structure that naturally incorporates spin while leveraging the intuition and efficiency benefits of crystallographic symmetry. For the generation of the \hamacro Hamiltonian, we provide an implementation into the Mathematica Software Package, GTPack. 
\end{abstract}
\maketitle
\section{Introduction}
Chemistry flourishes when synthetic efforts are driven by intuitive physical models that link structure to function. An iconic example from molecular chemistry is the control of the d-orbital manifold of transition metals via ligand coordination environment. In many cases, a surprising level of insight can be gained through the heuristic assumptions of a simplified symmetry and a basic electrostatic model (i.e. Crystal Field Theory; CFT) \cite{bethe1929termaufspaltung,danielsen1972quantum,mulak2000effective,liu2018symmetry}. The limitations of CFT are often cited as evidence for involvement of ligand orbitals and thus the necessity of an expanded basis allowing for mixing with ligand orbitals (Ligand Field Theory; LFT) \cite{griffith1957ligand}. Indeed, benchmark structure-function relationships such as the spectrochemical series and nephelauxetic effect \cite{nephelauxetic} require LFT insight \cite{jorgensen1966recent}. Rightly so, the influence of ligand orbitals is considered foundational knowledge for many chemists. Knowledge of orbital moment and the spin-orbit interaction, however, is much less pervasive, despite being vital to understanding of electronic structure, especially for the d-orbital manifold of many materials of high physical and technological interest. Even though CFT/LFT models do not explicitly treat angular momentum, basic qualitative predictions can be made in simple cases with only the Pauli exclusion principle as a guide (e.g. spin-only models). Caution in this respect is imperative, however, as the underlying physics of electronic spin is absent here, limiting insight in both scope and scale. 

Recognition of this issue goes back to the advent of quantum mechanics, and as disciplines became more specialized, approaches to the problem developed a dizzying array of formalisms. A common historic theme was that full spin quantum mechanical modeling or fitting was too computationally-intensive for practical use, and, in the end, unnecessary for empirical descriptions of most phenomena. For instance, exclusions of energy states outside of the range of interest vastly reduces the computational cost. Such simplifications trade physically meaningful operators for phenomenologically parameterized tensors to compensate for the simplified basis. Often referred to as spin Hamiltonians, \cite{li2021spin,rudowicz2015disentangling,mostafanejad2014basics,misra2003review}, such models are particularly effective in efficiently reproducing magnetometry and magnetic spectroscopy data of transition metal complexes when the ground state lacks orbital angular momentum. Developed largely in the context of Electron Paramagnetic Resonance (EPR) \cite{abragam2012electron}, spin Hamiltonians are characterized by a lack of orbital operators with compensation via a phenomenological symmetry-restricted Hamiltonian acting directly on functions of the spin operators. There are a number of common methods for perturbations involving the orbital operators where the symmetry of the local crystal field is expanded into operator equivalents following methods of Stevens \cite{stevens1952matrix,ryabov1999generation,rudowicz2004generalization} or Buckmaster-Smith-Thornley \cite{smith1966use}, for example. This approach has proven powerful and efficient, in particular in situations of well-defined symmetry and low anisotropy.

While effective in fitting complex spectral data, there are reasons to consider alternative approaches to the crystal field expansion. The first of these reasons being the lack of intrinsic physical meaning to the parameters. Without the ability to generate falsifiable hypotheses for predicted spin state structure from coordination environment, progress towards new, overarching models or technological goals is drastically hindered. Additionally, many of the most exciting systems of current interest are incompatible with the spin Hamiltonian model. In cases of low-lying excited states and significant orbital moment, for example, employing a spin Hamiltonian will provide little insight. As a final point, somewhat counter-intuitively, a more complete incorporation of the underlying spin and orbital moment can actually reduce the overall complexity despite an expansion of the Hilbert space. The spin symmetry lost in the truncated basis can often be leveraged by modern computational algorithms to considerable effect. In this work, we present a method for deploying point group representations of spin-symmetry called the \textit{\hamname} (\hamacro) method. \hamacro relies solely on symmetry constraints. By working in the full spin rotational group and subgroups thereof we are able to maintain the meaningful angular momentum operators with crystal field and spin orbit interactions arising via mixing terms that emerge from the symmetry restrictions.
\section{Method Overview}
 In the following section, we provide theoretical background outlining the \hamname approach. To efficiently evaluate the symmetry constraints imposed on the crystal field Hamiltonian, we make use of the Mathematica group theory package GTPack \cite{gtpack1,gtpack2}. In particular, we extend the functionality of GTPack to calculate representation matrices of O(3) and SU(2), as described throughout the main text and in the appendix. All examples presented here can be found in the Mathematica notebook format (Supporting information).
\begin{figure}[b!]
    \centering
    \includegraphics[height=8cm]{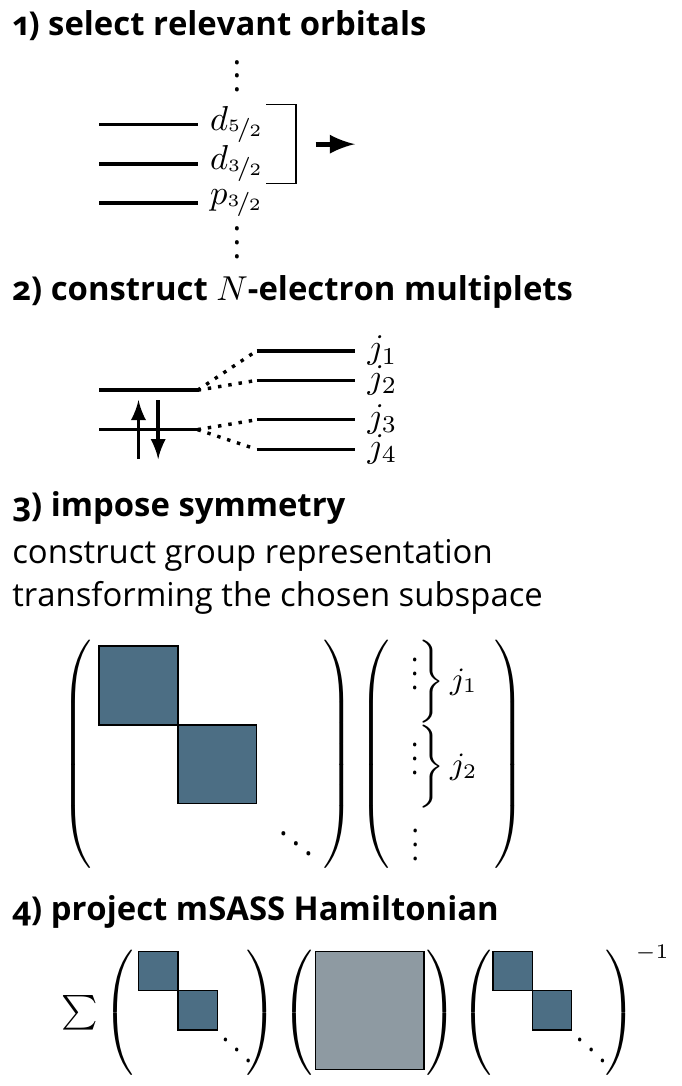}
    \caption{Outline of the \hamname method.}
    \label{method}
\end{figure}
\subsection{Spinor basis}
The \hamname (\hamacro) method is constructed in the spinor basis (Fig. \ref{method}). For total angular momentum $J$, a $(2J+1)$-dimensional basis is spanned by the functions $ \left|J m\right> $, with $m=-J,\dots,J$. For multiple values of $J$, the basis is extended accordingly, $\left\{\ket{J_1,J_1},\ket{J_1,J_1-1},\dots,\ket{J_1,-J_1},\ket{J_2,J_2},\ket{J_2,J_2-1},\dots,\ket{J_2,-J_2}\right\}$. 

\subsection{Multi-electron basis}
Many-particle basis functions are constructed in terms of direct products of single particle wave functions. For electrons, the wave function needs to satisfy fermionic statistics, i.e., being odd under particle-exchange. This constraint is satisfied by taking the alternating square of two representations, instead of the ordinary direct product \cite{james2001representations}. 

In the following we introduce the characters of $SU(2)$ representations and the construction of the alternating square. We start from rotation of angle $\phi$ about the arbitrary axis $\vec{n}$, $D^J(\vec{n},\phi)$, represented as 
\begin{equation}
 D^J(\vec{n},\phi) = D^J(\vec{n}',-\theta)D^J(\vec{e}_z,\phi)D^J(\vec{n}',\theta),
\end{equation}
with $\vec{n}'$ and $\theta$ being the respective axis and angle to transform the initial axis $\vec{n}$ to the $\vec{e}_z$ axis. Due to the cyclic permutation rule for the trace operation we obtain for the characters $\chi^J(\vec{n},\phi)$ and $\chi^J(\vec{e}_z,\phi)$, $\chi^J(\vec{n},\phi) = \operatorname{Tr} \left[D^J(\vec{n}',-\theta)D^J(\vec{e}_z,\phi)D^J(\vec{n}',\theta) \right] = \operatorname{Tr} \left[D^J(\vec{n}',\theta)D^J(\vec{n}',-\theta)D^J(\vec{e}_z,\phi) \right] = \operatorname{Tr} \left[D^J(\vec{e}_z,\phi) \right] = \chi^J(\vec{e}_z,\phi)$. Hence, without loss of generality, we focus on rotations about the $z$-axis in the following. 

Since, $D^J(\vec{e}_z,\phi) = \exp \left(-\mathrm{i} \op{J}_z \phi\right)$ and $\left< J,m |D^J(\vec{e}_z,\phi)| J' m' \right> = \delta_{mm'}\delta_{JJ'} \exp \left(-\mathrm{i} m' \phi\right)$ we obtain for the character
\begin{equation}
    \chi^J(\vec{e}_z,\phi) = \sum_{m=-J}^J e^{-\mathrm{i}m\phi}.
    \label{chars_SU2}
\end{equation}

We continue by considering the direct product of two $SU(2)$ representations, where we assume $J_1>J_2$. By calculating the product $\chi^{J_1}(\vec{e}_z,\phi)\chi^{J_2}(\vec{e}_z,\phi) = \chi^{J_1+J_2}(\vec{e}_z,\phi)+\chi^{J_1+J_2-1}(\vec{e}_z,\phi)+\dots+\chi^{J_1-J_2}(\vec{e}_z,\phi)$ we obtain
\begin{equation}
    D^{J_1}\otimes D^{J_2} \simeq D^{J_1+J_2}\oplus D^{J_1+J_2-1}\oplus \dots \oplus D^{J_1-J_2}.
\end{equation}
For $J_1 = J_2$, the representation matrices operate on the same finite vector space $\mathcal{V}^J$. The basis of the direct product space $\mathcal{V}^J \otimes \mathcal{V}^J$ can be constructed from products $v_i v_j$, where $v_i,v_j\in  \mathcal{V}^J$ are a basis of  $\mathcal{V}^J$. This direct product can be decomposed into a symmetric (bosonic) and an antisymmetric (fermionic) part,
\begin{equation}
    \mathcal{V}^J \otimes \mathcal{V}^J = S\left(\mathcal{V}^J \otimes \mathcal{V}^J\right) \oplus A\left(\mathcal{V}^J \otimes \mathcal{V}^J\right).
\end{equation}
The symmetric and antisymmetric parts of the decomposition are also known as the symmetric and antisymmetric square of the representation. Correspondingly, one obtains for the direct product representation $D^{J}\otimes D^{J} \simeq D^{S}\otimes D^{A}$ with $\chi^J(g)^2 = \chi^S(g) + \chi^A(g)$. The respective characters for the symmetric and the antisymmetric square are \cite{james2001representations},
\begin{align}
     \chi^S(g) &= \frac{1}{2}\left(\chi^J(g)^2 + \chi^J\left(g^2\right) \right), \\      \chi^A(g) &= \frac{1}{2}\left(\chi^J(g)^2 - \chi^J\left(g^2\right) \right) \label{chars_antisym},
\end{align} where $g\in\mathcal{G}$ is a symmetry element of the group $\mathcal{G}$. 

Finally, we give the \textit{decomposition of the antisymmetric square in terms of representations of $SU(2)$},  
 \begin{equation}
     A\left( D^{J}\otimes D^{J} \right)  \simeq D^{2 J-1}\oplus D^{2J-3}\oplus \dots .
     \label{antisym_square}
 \end{equation}
 The above equation can be obtained from applying \eqref{chars_SU2} to equation \eqref{chars_antisym}. 
 
 \textit{As an example}, we consider $J=\nicefrac{1}{2}$ and the well known direct product
 \begin{equation}
      D^{\nicefrac{1}{2}}\otimes D^{\nicefrac{1}{2}} \simeq D^{1}\oplus D^{0}.
 \end{equation}
 From equation \eqref{antisym_square} it follows that the antisymmetric part of the direct product is given by the spin-singlet $D^0$,
 \begin{equation}
     A\left( D^{\nicefrac{1}{2}}\otimes D^{\nicefrac{1}{2}}  \right)  \simeq D^{0} .
 \end{equation}
 The respective results for all direct products involving single values of $J$, relevant for $s-f$ electrons with and without spin orbit coupling are given in Table \ref{list_squares}.
 \begin{table}[b!]
 \caption{The antisymmetric squares for relevant direct products arising for $s-f$ electrons with and without spin-orbit coupling }
     \label{list_squares}
     \noindent\rule{\columnwidth}{0.4pt}
 \begin{align}
     A\left( D^{\nicefrac{1}{2}}\otimes D^{\nicefrac{1}{2}}  \right)  &\simeq D^{0} \notag \\
     A\left( D^{1}\otimes D^{1}  \right)  &\simeq D^{1} \notag \\
     A\left( D^{\nicefrac{3}{2}}\otimes D^{\nicefrac{3}{2}}  \right)  & \simeq D^{2} \oplus D^0 \notag \\
     A\left( D^{2}\otimes D^{2}  \right)  &\simeq D^{3} \oplus D^1 \notag \\
     A\left( D^{\nicefrac{5}{2}}\otimes D^{\nicefrac{5}{2}}  \right)  &\simeq D^{4} \oplus D^{2} \oplus D^0 \notag \\
     A\left( D^{3}\otimes D^{3}  \right)  &\simeq D^{5} \oplus D^3 \oplus D^1 \notag \\
     A\left( D^{\nicefrac{7}{2}}\otimes D^{\nicefrac{7}{2}}  \right)  &\simeq D^{6} \oplus D^{4} \oplus D^{2} \oplus D^0 \notag 
 \end{align}
  \noindent\rule{\columnwidth}{0.4pt}
 \end{table}
\subsection{Construction of the \hamacro Hamiltonian}
Consider a system with a basis belonging to several total angular momenta $J_i$, where each $J_i$ corresponds to a $(2J_i+1)$-dimensional subspace with basis function $u^{J_i}_m$ ($J_i \geq m \geq -J_i$). Furthermore, consider a crystal field with symmetry group $\mathcal{G}$. By definition, the transformation of $u^{J_i}_m$ under a symmetry element $g\in\mathcal{G}$ is given by
\begin{equation}
    g u^{J_i}_m = \sum_{m'=-J_i}^{J_i} D_{m'm}^{J_i}(g) u^{J_i}_{m'},
\end{equation}
with $D_{m'm}^{J_i}(g)$ denoting the matrix of the irreducible representation of SO(3) corresponding to $J_i$. In general, $D_{m'm}^{J_i}(g)$ form a reducible representation in $\mathcal{G}\subset SO(3)$. 

The total dimension of a Hamiltonian spanned by $(u^{J_1}_{J_i},u^{J_1}_{J_i-1},\dots u^{J_2}_{J_2},\dots)$ is given by
\begin{equation}
    d = \sum_{i} (2J_i+1).
\end{equation}
The corresponding Hamiltonian is a $d \times d$-dimensional Hermitian matrix~\footnote{We note that a Hermitian matrix was historically chosen to enforce real eigenvalues of the Hamiltonian. This condition can be lifted towards $\mathcal{P}\mathcal{T}$-symmetric matrices \cite{bender1999pt} which bring a new twist to quantum chemistry. Also, non-Hermitian matrices are meaningful, in the context of open quantum systems \cite{rotter2009non}.}.
\begin{equation}
    \mat{H} = \left(
    \begin{array}{cccc}
        h_{11} & h_{12} & \dots & h_{1d}  \\
        h^*_{12} & h_{22} & \dots & h_{2d} \\
        \vdots & \vdots & \ddots & \vdots\\
        h^*_{1d} & h^*_{2d} & \dots & h_{dd} 
    \end{array}
     \right)
\end{equation}
This matrix has to be invariant under all symmetry elements of $\mathcal{G}$, imposing
\begin{equation}
    \mat{H} = \sum_{g\in\mathcal{G}} \mat{D}(g)\mat{H}\mat{D}(g)^{-1},
    \label{symmetry_constraint}
\end{equation}
where $\mat{D}(g)$ denotes the super representation of $g$ incorporating all matrices $\mat{D}^{J_i}(g)$ as follows,
\begin{equation}
    \mat{D}(g) = \left(
    \begin{array}{cccc}
        \mat{D}^{J_1}(g) & 0 & \dots & 0  \\
        0 & \mat{D}^{J_2}(g) & \dots & 0 \\
        \vdots & \vdots & \ddots & \vdots\\
        0 & 0 & \dots & \mat{D}^{J_N}(g)
    \end{array}
     \right).
     \label{superrep}
\end{equation}
While $\mat{H}$ is diagonal in blocks belonging to different $J_i$, imposing \eqref{symmetry_constraint} for $\mathcal{G}\subset SO(3)$ separates $\mat{H}$ into blocks belonging to different irreducible representations of $\mathcal{G}$. The diagonal formulation of $\mat{H}$ in terms of $J_i$ represents the unrestricted rotational symmetry of the chosen angular momentum operators. If multiple types of angular momenta are used (e.g. $J = S + L$) intra-$J_i$ coupling terms naturally incorporate any associated interactions via symmetry (e.g. the spin-orbit interaction). When the further constraint of $\mathcal{G}\subset SO(3)$ is included, \eqref{symmetry_constraint} inter-$J_i$ terms arise naturally from the symmetry of $\mathcal{G}$ to account for the new crystal field interactions. Due to the symmetry constraints of real molecules, $D^{J_i}$ can now be reduced in $\mathcal{G}$ to give 
\begin{equation}
    D^{J_i} \sim n_1 \Gamma^1 \oplus n_2 \Gamma^2 \oplus \dots \oplus n_n \Gamma^n,
\end{equation}
where $\Gamma^i$ are the $n$ nonequivalent irreducible representations of $\mathcal{G}$. Hence, in the most general form, the \textit{\hamacro Hamiltonian} $\mat{H}$ is given by
\begin{equation}
    \mat{H} = \left(
    \begin{array}{ccccccc}
        \mat{H}_{\Gamma^{1,1}}  & \mat{h}_{1} & \dots & 0 & \dots & \mat{\Delta}_{1} & \dots \\
        \mat{h}^\dagger_{1} & \mat{H}_{\Gamma^{1,2}} & \dots & 0 & \dots & \mat{\Delta}_{2} & \dots \\
        \vdots & \vdots & \ddots & \vdots &  & \vdots &  \\
        0 & 0 & \dots & \mat{H}_{\Gamma^{i,\alpha}} &  \dots & 0 & \dots \\
        \vdots & \vdots &  & \vdots & \ddots & \vdots & \\
        \mat{\Delta}^\dagger_{1} & \mat{\Delta}^\dagger_{1} & \dots & 0 & \dots & \mat{H}_{\Gamma^{1,\beta}} & \dots\\
        \vdots & \vdots & & \vdots & & \vdots &
    \end{array}
    \right)
    \label{LEMONHam}
\end{equation}
Here $\mat{h}_i$ ($\mat{\Delta}_i$) are intra-hybridization (inter-hybridization) coupling terms belonging to equivalent irreducible representations $\Gamma^i$ of $\mathcal{G}$, but the same (different) $J$. The implementation of this formalism is shown below in several cases where pre-existing intuition will be helpful. In this work, we do not yet apply the method in complex or quantitative methods, instead focusing on establishing the groundwork for further development and implementation. As an aid in this process, the \hamacro Hamiltonian $\mat{H}$ can be deduced from evaluating equation \eqref{symmetry_constraint} using the Mathematica Software Package, GTPack. \cite{gtpack1, gtpack2} 

\section{Examples}
To evaluate equation \eqref{symmetry_constraint} for specific point group symmetries $g\in\mathcal{G}$, we implemented a new module \textit{GTAngularMomenumRep} into the Mathematica group theory package GTPack \cite{gtpack1,gtpack2}, calculating the $SU(2)$ and $SO(3)$ representation matrices $D^J(g)$. Full details of the implementation are given in the appendix and software documentation. Briefly, the starting point for implementation of orbital symmetry is often $O(3)$, the full roto-reflection group of $\mathbb{R}^3$ represented by orthonormal $3 \times 3$ matrices of $\det = \pm 1$). $O(3)$ is the direct product of SO(3) and the group $S_I=\left\{E, I\right\}$ ($E$ the identity, $I$ the inversion), $O(3)=SO(3)\times S_I$. The irreducible representations of O(3) come as even and odd variants. The physically relevant representations are even (odd) under inversion for even (odd) angular momenta $l$. Hence, for $d$-electrons, the corresponding irreducible representation of O(3) would be $D^2_g$.

\subsection{The $d^1$ configuration in an octahedral crystal field}
To introduce the methodology, We will build up the textbook example of a single $d$-electron in an octahedral, $O$, crystal field where spin is not a factor outside of the Pauli exclusion principle. In Section III.B, we will add detail through the introduction of spinor representations, constructing the fermionic picture in the presence of simultaneous symmetry restrictions of the spin-orbit and crystal field interactions. 

Considering only the orbital moment, the five-fold degenerate $d$-electron wave function in $O$-symmetry splits according to
\begin{equation}
    D^2 \simeq E \oplus T_{2}.
\end{equation}

Note that we choose $O$ for our analysis of the restrictions of the crystal field rather than the inversion symmetric $O_h$. The pure rotational octahedral group, $O$, provides a more straightforward example, as well as one of more general utility due to the ready extension to the tetrahedral case, $T$. 

In terms of real (tesseral) spherical harmonics $Y^l_m = \left< \vec{r} | l,m \right>$, the basis of the $5$-dimensional subspace belonging to $l=2$ is given by
\begin{equation}
    \left\{\ket{2,2}, \ket{2,1}, \ket{2,0}, \ket{2,-1} ,\ket{2,-2} \right\}.
\end{equation}
In this basis, the \hamacro Hamiltonian matrix is diagonal, giving rise to three-fold degenerate ($T_2$) and two-fold degenerate ($E$) electronic configurations,
\begin{equation}
\mat{H} = 
    \left( \begin{array}{ccccc}
        \epsilon_{E} & 0 & 0 & 0 & 0 \\
        0 & \epsilon_{T_2} & 0 & 0 & 0 \\
        0 & 0 & \epsilon_{E} & 0 & 0 \\
        0 & 0 & 0 & \epsilon_{T_2} & 0 \\
        0 & 0 & 0 & 0 & \epsilon_{T_2} 
    \end{array} \right).
    \label{Hamd1}
\end{equation}
To refer to the general form of the \hamacro Hamiltonian in \eqref{LEMONHam}, we note that the matrices $\mat{H}_E$ and $\mat{H}_{T_2}$ are diagonal, i.e., $\mat{H}_E = \operatorname{diag}\left(\epsilon_E,\epsilon_E\right)$ and $\mat{H}_{T_2} = \operatorname{diag}\left(\epsilon_{T_2},\epsilon_{T_2},\epsilon_{T_2}\right)$. As each representation only occurs once, the coupling matrices vanish, $\mat{h}_i = \mat{\Delta}_i = 0$.
The eigenfunctions of the \hamacro Hamiltonian for a $d^1$ configuration in an octahedral field can be constructed to transform as
\begin{align}
  \Psi_{T_2; 1} &\sim \left< \vec{r} | 2,1 \right> \sim  xz,\\
  \Psi_{T_2; 2} &\sim \left< \vec{r} | 2,-1 \right> \sim yz,\\
  \Psi_{T_2; 3} &\sim \left< \vec{r} | 2,-2 \right> \sim  xy,\\
  \Psi_{E; 1} &\sim \left< \vec{r} | 2,0 \right> \sim  z^2 - r^2,\\
  \Psi_{E; 2} &\sim \left< \vec{r} | 2,2 \right> \sim x^2 - y^2. 
\end{align}
Note that in the case of a single electron, these eigenfunctions have the mathematical form of the commonly used Cartesian d-orbital set. It is important, however, not to conflate the solutions to \eqref{Hamd1} (corresponding to possible configurations available to a $d^1$ electron count) with the concept of orbital energy levels. The corresponding Mathematica code can be found in the supplemental material. 

Throughout this work we focus mainly on the cubic symmetry $O$ as a representative example. Without modification, however, the algorithm used for $O$ may be applied to any point group symmetry. For example, the degeneracy of possible configurations for $d^1$ may be lowered from $O$ to $D_4$ by an axial Jahn-Teller distortion, leading to symmetry lowering according to:
\begin{align}
    E &\simeq A_1 \oplus B_1, \\
    T_2 &\simeq E + B_2.
\end{align}
In agreement with this decomposition, the \hamacro Hamiltonian \eqref{LEMONHam} is of the form
\begin{equation}
\mat{H} = 
    \left( \begin{array}{ccccc}
        \epsilon_{B_1} & 0 & 0 & 0 & 0 \\
        0 & \epsilon_{E} & 0 & 0 & 0 \\
        0 & 0 & \epsilon_{A_1} & 0 & 0 \\
        0 & 0 & 0 & \epsilon_{E} & 0 \\
        0 & 0 & 0 & 0 & \epsilon_{B_2} 
    \end{array} \right).
\end{equation}
The corresponding eigenfunctions of the \hamacro Hamiltonian for the $d^1$ configuration in square symmetry transform as follows
\begin{align}
  \Psi_{E; 1} &\sim \left< \vec{r} | 2,1 \right> \sim  xz,\\
  \Psi_{E; 2} &\sim \left< \vec{r} | 2,-1 \right> \sim yz, \\
  \Psi_{B_2} &\sim \left< \vec{r} | 2,-2 \right> \sim  xy,\\
  \Psi_{A_1} &\sim \left< \vec{r} | 2,0 \right> \sim  z^2 - r^2,\\
  \Psi_{B_1} &\sim \left< \vec{r} | 2,2 \right> \sim x^2 - y^2. 
\end{align}
A schematic of the hierarchical symmetry lowering of the $d$-state by $O$ and $D_4$ crystal fields is shown in Fig. \ref{d1:splitting}.  
\begin{figure}
    \centering
    \includegraphics[]{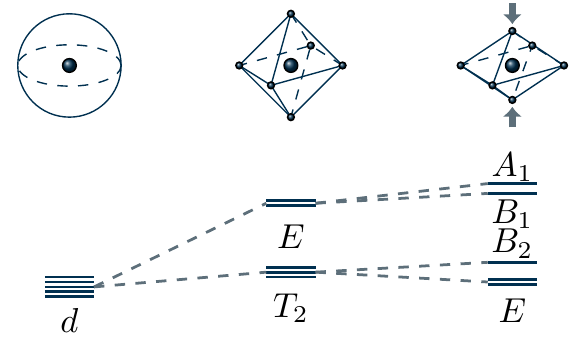}
    \caption{Splitting of a single $d$-electron in octahedral and square (bipyramidal) geometries, with symmetry groups $O$ and $D_4$. The axes represent relative energy changes vs distortion coordinates for the two symmetry transformations.}
    \label{d1:splitting}
\end{figure}

\subsection{The $d^1$ configuration in an octahedral crystal field with spin-orbit coupling}
Orbital symmetry has been arguably the greatest driver of intuitive models for chemical synthesis, properties, and reactivity in the era of modern quantum theory; spin-symmetry, on the other hand, has had only minor influence. While the reasons for this are complex and beyond the scope of this work, in many ways the triage of spin can be attributed to the search for configuration subspaces accessible by contemporary computational methods. These subspaces, though vastly reduced in dimensionality, lack the overarching symmetry restrictions of the original space. When applied to systems with non-trivial spin interactions, the reduced space may still effectively emulate the properties of the local region of the full space assuming the subspace remains largely orthogonal to the full configuration space (i.e. cases where interactions arising from spin comprise only very minor perturbations on the energy). In the framework of (16), $\mat{\Delta}_{i} \approx 0$.     

Because of the limitations of the restricted subspace to minor perturbations, the spin-orbit interaction is typically discussed in terms of two limits. In the first limit, the spin-orbit interaction dominates the crystal field, causing a splitting of the $d$-level into two levels with angular momenta $j=\nicefrac{3}{2}$ and $j=\nicefrac{5}{2}$. These levels then split in the presence of the crystal field. Continuing from our example of $O$ symmetry, the splitting follows
\begin{align}
    D^{\nicefrac{3}{2}} &\simeq F_{\nicefrac{3}{2}}, \\
    D^{\nicefrac{5}{2}} &\simeq E_{\nicefrac{5}{2}} \oplus F_{\nicefrac{3}{2}}.
\end{align}

The far more common case in valence electrons of chemical importance occurs when the crystal field dominates. Here, one can discuss the splitting of the $E$ and $T_2$ levels discussed previously, due to spin orbit interaction. In the case of the group $O$, these representations split as follows
\begin{align}
    E &\simeq F_{\nicefrac{3}{2}}, \\
    T_2 &\simeq E_{\nicefrac{5}{2}} \oplus F_{\nicefrac{3}{2}}.
\end{align}

\begin{figure}[]
    \centering
    \includegraphics[]{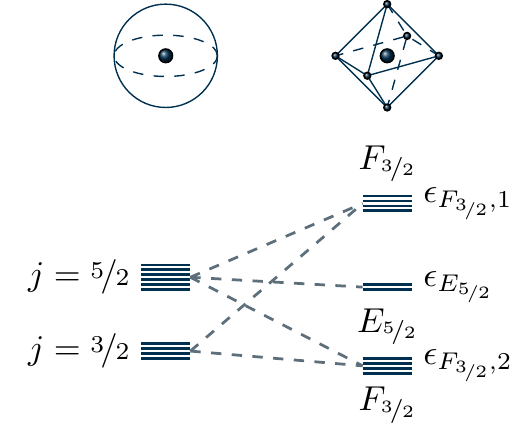}
    \caption{Splitting of a single $d$-electron in octahedral field with symmetry group $O$ and spin-orbit interaction. In the presence of both, the $j=\nicefrac{3}{2}$ and $j=\nicefrac{5}{2}$ multiplets, states belonging to the irreducible representation $F_{\nicefrac{3}{2}}$ mix, an effect which is usually neglected in the strong spin-orbit limit.}
    \label{d1:splitting}
\end{figure}

In general, however, the two levels transforming as $F_{\nicefrac{3}{2}}$ interact. To obtain the corresponding \hamacro Hamiltonian, we constrain the problem to the 10-dimensional subspace containing spinor functions belonging to the angular momenta $j_1 = \nicefrac{3}{2}$ and $j_2=\nicefrac{5}{2}$,
\begin{equation}
    \left\{\ket{\nicefrac{3}{2},\nicefrac{3}{2}}, \dots \ket{\nicefrac{3}{2},-\nicefrac{3}{2}}, \ket{\nicefrac{5}{2},\nicefrac{5}{2}},\dots,\ket{\nicefrac{5}{2},-\nicefrac{5}{2}}\right\}.
    \label{basis_d1_SOC}
\end{equation}
The corresponding super representation \eqref{superrep} is given by the Clebsch-Gordan sum
\begin{equation}
    D = D^{\nicefrac{3}{2}} \oplus D^{\nicefrac{5}{2}}.
\end{equation}
As there are two equivalent representations ($F_{\nicefrac{3}{2}}$) and one additional representation ($E_{\nicefrac{5}{2}}$), we expect four free parameters in our theory, three determining the three levels, plus one parameter determining the interaction between the levels belonging to the equivalent representation $F_{\nicefrac{3}{2}}$. Evaluating \eqref{symmetry_constraint}, our algorithm gives the following \hamacro Hamiltonian matrix for $d^1$ configurations under the full spin-orbit symmetry, with the parameters $\lambda_1,\dots,\lambda_4$,
\begin{widetext}
\begin{equation}
\mat{H} = 
\left(
\begin{array}{cccccccccc}
\lambda_1& 0 & 0 & 0 & 0 &\lambda_4& 0 & 0 & 0 & \sqrt{5}\lambda_4\\
 0 &\lambda_1& 0 & 0 & 0 & 0 & -\sqrt{6}\lambda_4& 0 & 0 & 0 \\
 0 & 0 &\lambda_1& 0 & 0 & 0 & 0 & \sqrt{6}\lambda_4& 0 & 0 \\
 0 & 0 & 0 &\lambda_1& -\sqrt{5}\lambda_4& 0 & 0 & 0 & -\lambda_4 & 0 \\
 0 & 0 & 0 & -\sqrt{5} \lambda_4^* &\lambda_2& 0 & 0 & 0 &\lambda_3& 0 \\
 \lambda_4^* & 0 & 0 & 0 & 0 & \lambda_2-\frac{4 \lambda_3}{\sqrt{5}} & 0 & 0 & 0 &\lambda_3\\
 0 & -\sqrt{6} \lambda_4^* & 0 & 0 & 0 & 0 & \lambda_2+\frac{\lambda_3}{\sqrt{5}} & 0 & 0 & 0 \\
 0 & 0 & \sqrt{6} \lambda_4^* & 0 & 0 & 0 & 0 & \lambda_2+\frac{\lambda_3}{\sqrt{5}} & 0 & 0 \\
 0 & 0 & 0 & -\lambda_4^* & \lambda_3 & 0 & 0 & 0 & \lambda_2-\frac{4 \lambda_3}{\sqrt{5}} & 0 \\
 \sqrt{5} \lambda_4^* & 0 & 0 & 0 & 0 & \lambda_3 & 0 & 0 & 0 &\lambda_2\\
\end{array}
\right).
\end{equation}
\end{widetext}
By diagonalizing the matrix, one of the four parameters can be seen to represent the absolute energy scale and set to zero or an arbitrary offset. 
For real-valued parameters, the resulting energies for each term can be expressed as follows,
\begin{align}
    \epsilon_{E_{\nicefrac{5}{2}}} &= \lambda_2 - \sqrt{5} \lambda_3 \\
    \epsilon_{F_{\nicefrac{3}{2}},1} &= a - b,\\
    \epsilon_{F_{\nicefrac{3}{2}},2} &= a + b,
\end{align}
with
\begin{align}
    a &= \frac{1}{2}\left(\lambda_1+ \lambda_2+\frac{1}{\sqrt{5}} \lambda_3\right), \\
    b &= \sqrt{a^2-\lambda_1 \left(a - \lambda_1\right) +6 \lambda_4^2 }.
\end{align}
The corresponding eigenvectors are parameter dependent and are expressed as a linear combination of the basis functions given in \eqref{basis_d1_SOC}. As the eigenvectors transform as irreducible representations retaining the full spinor symmetry, they can be used to form a symmetry-restricted model of the full configuration space as a function of the full parameter space. Such models can provide useful insight on the often opaque intermediate configuration space away from the strong crystal field or spin-orbit limits. Additionally, the use of the symmetry restricted full configuration space allows for a more physically-meaningful model for fitting spin-dependent data. This flows from the fact that a given physical measurement often represents only a fraction of the energy range needed to parameterize a complex electronic structure. If the model is constructed for the energy range of the measurement instead of the energy range of the physical parameters, the relationship between the fitted parameters and the ``true'' ones is ill-defined and generally not useful for comparison across different systems.     

\subsection{Continuous symmetry parameterization of multielectronic configurations}
In the last example, we provide a simple extension to the multielectron case: a $p^4$ configuration within an octahedral field. This case extends the coordination chemistry examples used so far and serves as a demonstration of the versatility offered by taking a wider view of electronic structure problems. Configurations of several $p$-electrons occur e.g., in color centers or vacancies, and play an important role in $d^0$-magnetism \cite{coey2019magnetism,esquinazi2013defect} and NV centers \cite{Gali2017,Gali2008,Manson2006,Lenef1996}. For simplicity, we focus on two electrons in the $p_{\nicefrac{3}{2}}$-state, assuming that the $p_{\nicefrac{1}{2}}$ state is fully occupied and sufficiently separated in energy from $p_{\nicefrac{3}{2}}$. The corresponding alternating square from Tab. \ref{list_squares} is
\begin{equation}
    A\left( D^{\nicefrac{3}{2}}\otimes D^{\nicefrac{3}{2}}  \right)  \simeq D^{2} \oplus D^0.
\end{equation}
As before, we choose a basis in terms of angular momentum states. However, here we need to consider 2-electron wave functions, 
\begin{equation}
     \ket{J,M_J;j_1j_2} = \left\{\ket{2,2;\nicefrac{3}{2}\nicefrac{3}{2}},\dots,\{\ket{2,-2;\nicefrac{3}{2}\nicefrac{3}{2}}, \ket{0,0;\nicefrac{3}{2}\nicefrac{3}{2}} \right\}.
    \label{2e_basis}
\end{equation}
which are constructed using Clebsch-Gordan coefficients and the addition of angular momenta,
\begin{align}
\ket{2,2;\nicefrac{3}{2}\nicefrac{3}{2}} &= \frac{1}{\sqrt{2}}\left(\ket{\nicefrac{3}{2} \nicefrac{3}{2}} \ket{\nicefrac{3}{2} \nicefrac{1}{2}} - \ket{\nicefrac{3}{2} \nicefrac{1}{2}} \ket{\nicefrac{3}{2} \nicefrac{3}{2}}  \right), \\
\ket{2,1;\nicefrac{3}{2}\nicefrac{3}{2}} &= \frac{1}{\sqrt{2}}\left(\ket{\nicefrac{3}{2} \nicefrac{3}{2}} \ket{\nicefrac{3}{2} -\nicefrac{1}{2}} - \ket{\nicefrac{3}{2} -\nicefrac{1}{2}} \ket{\nicefrac{3}{2} \nicefrac{3}{2}}  \right), \\
\ket{2,0;\nicefrac{3}{2}\nicefrac{3}{2}} &= \frac{1}{2}\left(
    \ket{\nicefrac{3}{2} \nicefrac{3}{2}} \ket{\nicefrac{3}{2} -\nicefrac{3}{2}} + \ket{\nicefrac{3}{2} \nicefrac{1}{2}} \ket{\nicefrac{3}{2} -\nicefrac{1}{2}} \right. \notag\\&- \left.
    \ket{\nicefrac{3}{2} -\nicefrac{3}{2}} \ket{\nicefrac{3}{2} \nicefrac{3}{2}} - \ket{\nicefrac{3}{2} -\nicefrac{1}{2}} \ket{\nicefrac{3}{2} \nicefrac{1}{2}}\right), \\
\ket{2,-1;\nicefrac{3}{2}\nicefrac{3}{2}} &= \frac{1}{\sqrt{2}}\left(\ket{\nicefrac{3}{2} \nicefrac{1}{2}} \ket{\nicefrac{3}{2} -\nicefrac{3}{2}} - \ket{\nicefrac{3}{2} -\nicefrac{3}{2}} \ket{\nicefrac{3}{2} \nicefrac{1}{2}}  \right), \\
\ket{2,-2;\nicefrac{3}{2}\nicefrac{3}{2}} &= \frac{1}{\sqrt{2}}\left(\ket{\nicefrac{3}{2} -\nicefrac{1}{2}} \ket{\nicefrac{3}{2} -\nicefrac{3}{2}} \right. \notag\\&\qquad- \left. \ket{\nicefrac{3}{2} -\nicefrac{3}{2}} \ket{\nicefrac{3}{2} -\nicefrac{1}{2}}  \right), \\
\ket{0,0;\nicefrac{3}{2}\nicefrac{3}{2}} &= \frac{1}{2}\left(
    \ket{\nicefrac{3}{2} \nicefrac{3}{2}} \ket{\nicefrac{3}{2} -\nicefrac{3}{2}} - \ket{\nicefrac{3}{2} \nicefrac{1}{2}} \ket{\nicefrac{3}{2} -\nicefrac{1}{2}} \right. \notag\\& - \left.
    \ket{\nicefrac{3}{2} -\nicefrac{3}{2}} \ket{\nicefrac{3}{2} \nicefrac{3}{2}} + \ket{\nicefrac{3}{2} -\nicefrac{1}{2}} \ket{\nicefrac{3}{2} \nicefrac{1}{2}}\right).
\end{align}
In $O$ symmetry, the representations $D^2$ and $D^0$ are decomposed as before,
\begin{equation}
    D^{2} \simeq E \oplus T_2,\qquad D^{0}\simeq A_1.
\end{equation}
In contrast to the single $d$-electron Hamiltonian in \eqref{Hamd1}, we choose a complex valued representation according to our basis in \eqref{2e_basis}. Applying equation \eqref{symmetry_constraint} as before, we obtain a \hamacro Hamiltonian with 3 free parameters,
\begin{equation}
    \mat{H} = 
    \left( \begin{array}{cccccc}
        \lambda_1 & 0 & 0 & 0 & \lambda_3 & 0\\
        0 & \lambda_1 - \lambda_3 & 0 & 0 & 0 & 0\\
        0 & 0 & \lambda_1 + \lambda_3 & 0 & 0 & 0\\
        0 & 0 & 0 & \lambda_1 - \lambda_3 & 0 & 0\\
        \lambda_3 & 0 & 0 & 0 & \lambda_1 & 0 \\
        0 & 0 & 0 & 0 & 0 & \lambda_2
    \end{array} \right).
\end{equation}
\begin{figure}[b!]
    \centering
    \includegraphics[]{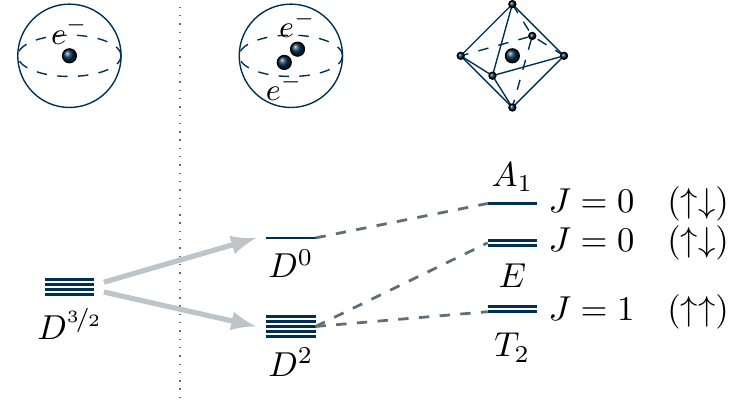}
    \caption{Splitting of energy levels for two $p$-electrons in an octahedral field.}
    \label{splitting-p2}
\end{figure}
The eigenvalues of this Hamiltonian are the level energies
\begin{align}
    \epsilon_{T_2} &= \lambda_1 - \lambda_3, \\
    \epsilon_{E} &= \lambda_1 + \lambda_3, \\
    \epsilon_{A_1} &= \lambda_2.
\end{align}
The eigenvectors of the Hamiltonian in terms of the respective two-electron basis functions are given by
\begin{align}
  \Psi_{T_2; 1} &\sim \ket{2,1;\nicefrac{3}{2}\nicefrac{3}{2}}, \\
  \Psi_{T_2; 2} &\sim \frac{1}{\sqrt{2}}\left( \ket{2,-2;\nicefrac{3}{2}\nicefrac{3}{2}} - \ket{2,2;\nicefrac{3}{2}\nicefrac{3}{2}} \right),\\
  \Psi_{T_2; 3} &\sim \ket{2,-1;\nicefrac{3}{2}\nicefrac{3}{2}}, \\
  \Psi_{E; 1} &\sim \ket{2,0;\nicefrac{3}{2}\nicefrac{3}{2}}, \\
  \Psi_{E; 2} &\sim \frac{1}{\sqrt{2}}\left( \ket{2,-2;\nicefrac{3}{2}\nicefrac{3}{2}} + \ket{2,2;\nicefrac{3}{2}\nicefrac{3}{2}} \right), \\
  \Psi_{A_1} &\sim \ket{0,0;\nicefrac{3}{2}\nicefrac{3}{2}}. 
\end{align}
As can be seen, the $T_2$ state corresponds to a spin-triplet with $J=1$, whereas the levels $E$ and $A_1$ give rise to spin singlets with $J=0$. The result is summarized in Fig. \ref{splitting-p2}.

\section{Conclusion}
Ideally, computational modeling, synthesis, and physical characterization of spin physics in molecular species would all fall under an overarching, universal framework robust enough to allow scientists to communicate through everything from back-of-the-envelope to supercomputer computations.

The \hamname (\hamacro) method allows for a generalization of the crystal field theory for calculating atomic spectra, taking into account the mixing of levels with different angular momenta, when exposed to a symmetry breaking molecular environment. The \hamacro Hamiltonian is constructed by, first, fixing a subspace consisting of several angular momentum multiplets belonging to a central atom or cluster, and, second, imposing symmetry constraints of the molecular environment. We outlined the approach on three simple examples, the $d^1$ configuration in octahedral and square symmetry, the $d^1$ configuration in octahedral symmetry with spin-orbit interaction, and, the $p^2$ configuration in an octahedral field. 

By way of introduction, we have provided a case for the need, theoretical basis, and simple implementation of the method, leaving detailed exploration of specific applications and fitting to experiments for subsequent work. These works will include, for example, symmetry-adapted splitting of states in a magnetic field.

We close by providing one more advantage of the \hamacro method, compared to conventional crystal field theory. Note, that evaluating \eqref{symmetry_constraint} is generic. By lifting the Hermiticity of the \hamacro Hamiltonian, the method can be generalized towards open quantum systems, i.e., quantum systems interacting with a thermal bath \cite{koch2016controlling,rotter2015review,rotter2009non}. In such systems, electrons occupying certain levels decay in time, which is a natural description for excited local electrons. A non-Hermitian \hamacro Hamiltonian contains the respective decay channels in terms of generally complex parameters. As such, it might play a powerful role, e.g., in optimizing the coherence time of molecular qubits \cite{gaita2019molecular} by symmetry principles. 

\section*{Author Contributions}
All authors contributed equally. RMG implemented the necessary algorithms into the Mathematica group theory package GTPack.

\section*{Conflicts of interest}
There are no conflicts to declare.

\section*{Acknowledgements}
We thank Wolfram Hergert for helpful discussions. RMG acknowledges support from Chalmers University of Technology. JDR acknowledges the National Science Foundation Division of Chemistry CHE-2154830.

\section*{Appendix}

\subsection*{Irreducible representation of SU(2)}
In the process of the present work, a new module \textit{GTAngularMomenumRep} was implemented into the Mathematica group theory package GTPack \cite{gtpack1,gtpack2}, calculating the $SU(2)$ and $SO(3)$ representation matrices. We follow the approach derived by Altmann \cite{altmann2005rotations}. Note that the $SU(2)$ matrices represent the $j=\nicefrac{1}{2}$ representation of $SU(2)$. We obtain all other representations by constructing corresponding tensor products as described below. As $SU(2)$ is the double cover of $SO(3)$, all $SO(3)$ representations are obtained in terms of integer valued $j$. The $j=\nicefrac{1}{2}$ representation is spanned by two basis functions $\mu_{-\nicefrac{1}{2}}$ and $\mu_{\nicefrac{1}{2}}$, such that
\begin{equation}
    \op{g}\mu_i = \sum_{j=-\nicefrac{1}{2}}^{\nicefrac{1}{2}} \mat{R}_{ji}(g) \mu_j.
\end{equation}
The basis can be used to construct tensors $\mu_{ijk\dots} = \mu_i\mu_j\mu_k\dots$, transforming as
\begin{equation}
    \op{g}\mu_{ijk\dots} = \sum_{IJK\dots} \mat{R}_{Ii}(g)\mat{R}_{Jj}(g)\mat{R}_{Kk}(g)\dots \mu_{ijk\dots}.
\end{equation}
We constrain ourselves to focus on the totally symmetric tensors having the property $\mu_{i\dots mn\dots} = \mu_{i\dots nm\dots}$, i.e., they are symmetric with respect to pairwise permutation of indices. Hence, we can order indices such that all values $-\nicefrac{1}{2}$ are listed on the left and all $\nicefrac{1}{2}$ on the right. This construction can be abbreviated by a function $v^j_m$ as follows
\begin{equation}
    v_m^j = \mu_{\underbrace{-\nicefrac{1}{2} -\nicefrac{1}{2} \dots}_{j-m \text{~times}} \underbrace{\nicefrac{1}{2}\,\nicefrac{1}{2} \dots}_{j+m \text{~times}}} = \mu^{j-m}_{-\nicefrac{1}{2}}\mu^{j+m}_{\nicefrac{1}{2}}.
\end{equation}
Acting on $v_m^j$ with the angular momentum operator $\op{J}_z$ gives
\begin{equation}
    \op{J}_z v_m^j = \left(-(j-m)\frac{1}{2}+(j+m)\frac{1}{2}\right)v_m^j = m v_m^j.
\end{equation}
Note that $\op{J}_z$ acts like a differential operator, i.e., $\op{J}_z \mu_{m_1}\mu_{m_2} = \left(\op{J}_z \mu_{m_1}\right) \mu_{m_2} +  \mu_{m_1}\left(\op{J}_z\mu_{m_2}\right)$. Normalizing $v_m^j$ leads to the final form
\begin{equation}
 v_m^j = \left((j+m)!(j-m)!\right)^{-\nicefrac{1}{2}}\mu^{j-m}_{-\nicefrac{1}{2}}\mu^{j+m}_{\nicefrac{1}{2}}.
 \label{jbasis}
\end{equation}
The representation matrices for higher $j$ can now be deduced by acting on the normalized $v_m^j$. A general transformation using SU(2) matrices for $j=\nicefrac{1}{2}$ can be written as
\begin{equation}
    \left(\begin{array}{c}
         \mu_{-\nicefrac{1}{2}}  \\
         \mu_{\nicefrac{1}{2}} 
    \end{array}\right)
    \left(\begin{array}{cc}
         a & b  \\
         -b^* & a^*
    \end{array}\right)
    =  \left(\begin{array}{c}
         a \mu_{-\nicefrac{1}{2}} - b* \mu_{\nicefrac{1}{2}}   \\
         b\mu_{-\nicefrac{1}{2}}  + a^* \mu_{\nicefrac{1}{2}} 
    \end{array}\right).
\end{equation}
Using \eqref{jbasis} we therefore obtain
\begin{multline}
    \op{g}v_m^j = \left((j+m)!(j-m)!\right)^{-\nicefrac{1}{2}}\left(a \mu_{-\nicefrac{1}{2}} - b* \mu_{\nicefrac{1}{2}}\right)^{j-m}\times \\ \times \left( b\mu_{-\nicefrac{1}{2}}  + a^* \mu_{\nicefrac{1}{2}} \right)^{j+m}_{\nicefrac{1}{2}}.
\end{multline}
This expression can be expanded using the binomial theorem, giving
\begin{multline}
    \op{g}v_m^j = \left((j+m)!(j-m)!\right)^{-\nicefrac{1}{2}}\times \\ \times \sum_{k=0}^{j+m}\sum_{\kappa=0}^{j-m}\frac{(j+m)!(j-m)!}{(j+m-k)!(j-m-\kappa)!k!\kappa!} \times \\
    \times a^{j+m-k}(a^*)^\kappa b^{j-m-\kappa}(b^*)^k \mu^{2j-k-\kappa}_{-\nicefrac{1}{2}} \mu^{k+\kappa}_{\nicefrac{1}{2}}.
    \label{jmatprior}
\end{multline}
Rewriting the indices as $2j-k-\kappa = j+m'$ and $k+\kappa=j-m'$ and noting that $j-k\leq m' \leq m-k$, leads to the final equation
\begin{multline}
    \op{g}v_m^j \\ = c^{j}_{mm'} \sum_{m'=j-k}^{m-k}\sum_{k=0}^{j+m} a^{j + m - k}(-b^*)^k b^{m' - m + k} (a^*)^{j-m' - k} v_{m'}^j,
     \label{jmat}
\end{multline}
with
\begin{equation}
   c^{j}_{mm'} =   \left((j+m)!(j-m)!(j+m')!(j-m')!\right)^{\nicefrac{1}{2}}.
\end{equation}
Equation \eqref{jmat} is used to determine the representation matrices $\mat{D}^j(g)$, which satisfy
\begin{equation}
    \op{g}v_m^j = \sum_{m'=-j}^j D_{m'm}^j(g) v_{m'}^j.
\end{equation}
The limits where either $a=0$ or $b=0$ can be derived straightforwardly from \eqref{jmatprior} without  using the binomial series.

\subsection*{Character Tables}
The character tables are calculated using a Burnside algorithm, implemented in GTPack. We choose the Mulliken notation for the irreducible representations \cite{Mulliken1955,Mulliken1956}, as tabulated by Altmann and Herzig \cite{altmann1994}. Representations under the separation are double group representations (see \cite{gtpack2} for more information).

\begin{table}[h!]
\caption{Top: character table of $O$. Bottom: character table of $D_4$.}
    \label{cts}
    \small
    \centering
    \begin{tabular}{cccccc}
    \hline\hline
         & E & $3C_2$ & $8C_3$ & $6C_4$ & $6C_2'$ \\
         \hline
        $A_1$ & 1 & 1 & 1 & 1 & 1 \\ 
        $A_2$ & 1 & 1 & 1 & -1 & -1 \\ 
        $E$ & 2 & 2 & -1 & 0 & 0 \\ 
        $T_1$ & 3 & -1 & 0 & 1 & -1 \\ 
        $T_2$ & 3 & -1 & 0 & -1 & 1 \\ 
\hline
        $E_{\nicefrac{1}{2}}$ & 2 & 0 & 1 & $\sqrt{2}$ & 0 \\ 
        $E_{\nicefrac{5}{2}}$ & 2 & 0 & 1 & $-\sqrt{2}$ & 0 \\ 
        $F_{\nicefrac{3}{2}}$ & 4 & 0 & -1 & 0 & 0 \\ 
\hline\hline
    \end{tabular}
    \hspace{0.4cm}    
    \raisebox{0.2cm}{
    \begin{tabular}{cccccc}
    \hline\hline
         & E & $2C_4$ & $C_2$ & $2C_2'$ & $2C_2''$ \\
         \hline
        $A_1$ & 1 & 1 & 1 & 1 & 1 \\ 
        $A_2$ & 1 & 1 & 1 & -1 & -1 \\ 
        $B_1$ & 1 & -1 & 1 & 1 & -1 \\ 
        $B_2$ & 1 & -1 & 1 & -1 & 1 \\ 
        $E$ & 2 & 0 & -2 & 0 & 0 \\ 
\hline
        $E_{\nicefrac{1}{2}}$ & 2 & $\sqrt{2}$ & 0 & 0 & 0 \\ 
        $E_{\nicefrac{3}{2}}$ & 2 & -$\sqrt{2}$ & 0 & 0 & 0 \\ 
        \hline\hline
    \end{tabular}}
\end{table}


%

\end{document}